\documentclass[doublecol]{epl2}
\usepackage{amsmath}
\usepackage{amssymb}
\usepackage{amsfonts}
\usepackage{amsthm}
\usepackage{graphics}
\usepackage{soul}

\newcommand\Rh{{\text{Rh}}}
\newcommand\Rey{{\text{Re}}}
\newcommand\Ro{{\text{Ro}}}

\def\mbfx{\mathbf{x}}

\def\mbfu{\mathbf{u}}
\def\mbfk{\mathbf{k}}
\def\mbfe{\mathbf{e}}
\def\mbff{\mathbf{f}}

\begin{document}

\title{Nonequilibrium Continuous Transition in a Fast Rotating Turbulence}
\author{Chandra Shekhar Lohani\inst{1}\thanks{E-mail: \email{cslohani25@gmail.com}}, Suraj Kumar Nayak\inst{1}\thanks{E-mail:  \email{surajkumarnayak96@gmail.com}}, Kannabiran Seshasayanan\inst{2} \thanks{E-mail: \email{s.kannabiran@gmail.com}} \and Vishwanath Shukla\inst{1}\thanks{E-mail: \email{research.vishwanath@gmail.com}}}
\shortauthor{Chandra Shekhar Lohani \etal}
\institute{
\inst{1} Department of Physics, Indian Institute of Technology Kharagpur, Kharagpur - 721 302, India\\
\inst{2} Department of Applied Mechanics \& Biomedical Engineering, Indian Institute of Technology Madras, 600036, India
}

\abstract{We study the saturation of three-dimensional unstable perturbations on a fast rotating turbulent flow using direct numerical simulations (DNSs). Under the effect of Kolmogorov forcing, a transition between states dominated by coherent two-dimensional modes to states with three-dimensional variations (quasi-two-dimensional) is observed as we change the global rotation rate. We find this akin to a critical phenomenon, wherein the order parameter scales with the distance to the critical point raised to an exponent. The exponent itself deviates from the predicted mean field value. Also, the nature of the fluctuations of the order parameter near the critical point indicate the presence of on-off intermittency. The critical rotation rate at which the transition occurs exhibits a linear scaling behaviour with the forcing wave number. A reduced model based on linear stability analysis is used to find the linear threshold estimates; we find these to be in good agreement with the 3D nonlinear DNS results.}

\maketitle

\section{Introduction}

Dimensionality plays an important role in determining the behaviour of physical systems. This is true for both quantum and classical fluids, irrespective of whether they are in equilibrium or out-of-equilibrium~\cite{muller2020abrupt,ShuklaNoneqBEC22}.  When driven out-of-equilibrium to a turbulent state, the flow organizes in different ways depending on the dimensionality and exhibits remarkably different spectral transport~\cite{verma2019energy}. A fully developed three-dimensional (3D) turbulence is characterized by the presence of an energy cascade towards smaller length scales, spatio-temporal intermittency, and an anomalous energy dissipation~\cite{frisch1995turbulence}, the latter being absent in two-dimensional (2D) turbulence~\cite{boffetta2012two,pandit2017overview}. 

In recent years, there has been a great interest in elucidating the emergent flow properties when the dimensionality is changed from 2D to 3D or vice versa. The effective dimensionality can be changed by the application of rotation~\cite{smith1996crossover,deusebio2014dimensional, buzzicotti2018inverse,pestana2019regime,sharma2019anisotropic,van2020critical}, stratification~\cite{marino2013inverse,oks2017inverse} and, of course, by changing the geometrical aspect ratio of the flow domain or confinement~\cite{celani2010turbulence, benavides2017critical,van2019condensates, van2019rare,alexakis2018cascades}. This change in effective dimensionality can cause the flow to transition from one non-equilibrium state to another with distinct features and can serve as an example of a non-equilibrium phase transition or bifurcations over a turbulent background~\cite{shukla2016statistical}, which have major consequences for the astrophysical, planetary, geophysical, and industrial flows. During these transitions, the behavior of characteristic features, such as the stability of 2D inverse cascade against 3D motions, is strongly influenced by the rotation rate and/or the geometrical aspect ratio. Hence, a key question is to understand the nature of the transition. 

Recent studies have shown that it is often possible to identify a suitable order parameter associated with these nonequilibrium transitions. When the effective dimensionality, a control parameter, is varied~\cite{van2024phase}, these transitions can be akin to either an abrupt~\cite{alexakis2015rotating, yokoyama2017hysteretic} or a continuous transition~\cite{seshasayanan2018condensates, van2020critical, alexakis2021symmetry}. The characterization of associated critical exponents is an open question. For example, a comprehensive understanding of the following still evades us: What are the universality classes, critical exponents, the deviation of the observed exponents from mean-field values, and the influence of intermittency, if any. In Refs.~\cite{benavides2017critical,alexakis2021symmetry} intermittency was observed close to the critical point. In the subcritical cases, the transition from quasi-2D to 3D or vice-versa was found to be exhibit bistability between two nonequilibrium states~\cite{yokoyama2017hysteretic,van2019condensates,ravelet2004multistability}. 

In this letter, we elucidate the transition from a 2D to a quasi-2D regime of a fast rotating incompressible fluid flow in a periodic box. This quasi-2D regime consists of coherent columnar vortices and sustained 3D perturbations. We study the full nonlinear problem using 3D direct numerical simulations (DNSs) in the presence of a large scale friction. We illustrate the behavior of the system close to the critical point by quantifying the  critical exponent, examining the flow structures and the energy cascades. We characterize the instability mechanism near the bifurcation, which is consistent with predictions of a reduced model~\cite{lohani2024effect}; moreover, it shows that a linear stability analysis (LSA) on a turbulent background is able to capture the essential features of this transition.

\begin{figure*}
	\centering
	\includegraphics[width=0.48\linewidth]{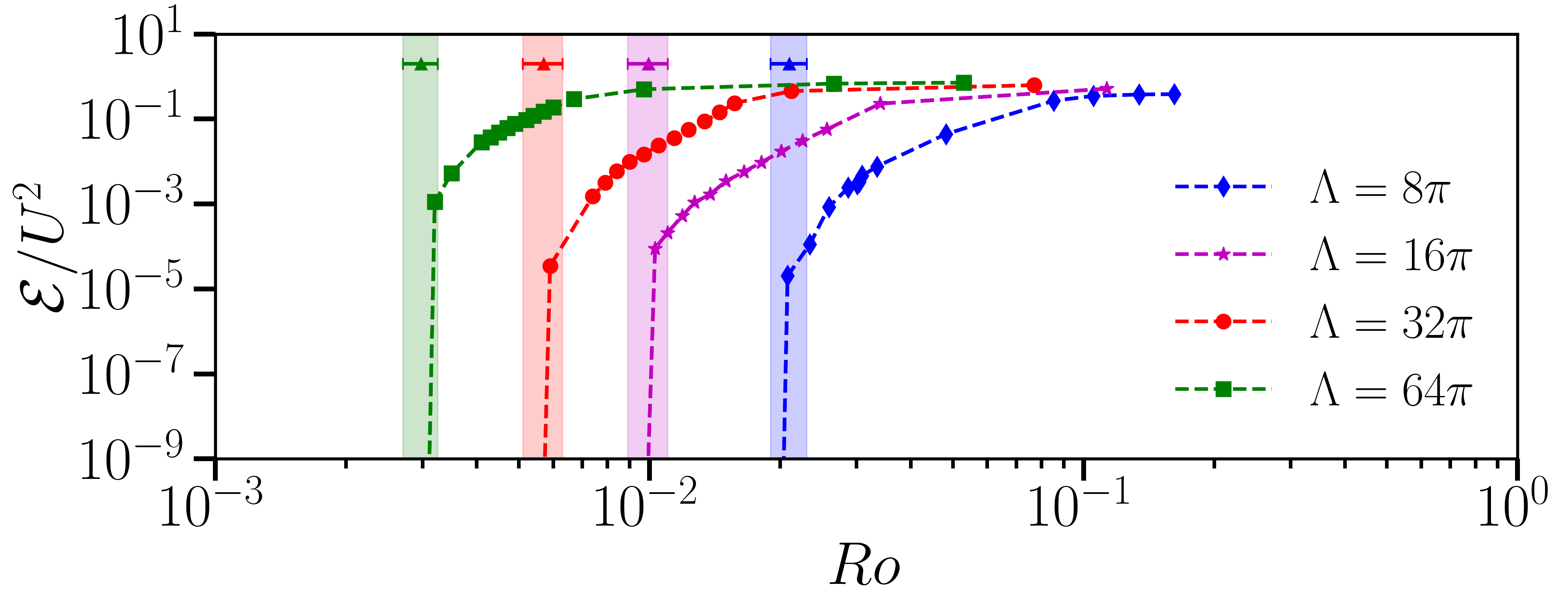}
	\put(-208,80){{(a)}}
	\includegraphics[width=0.48\linewidth]{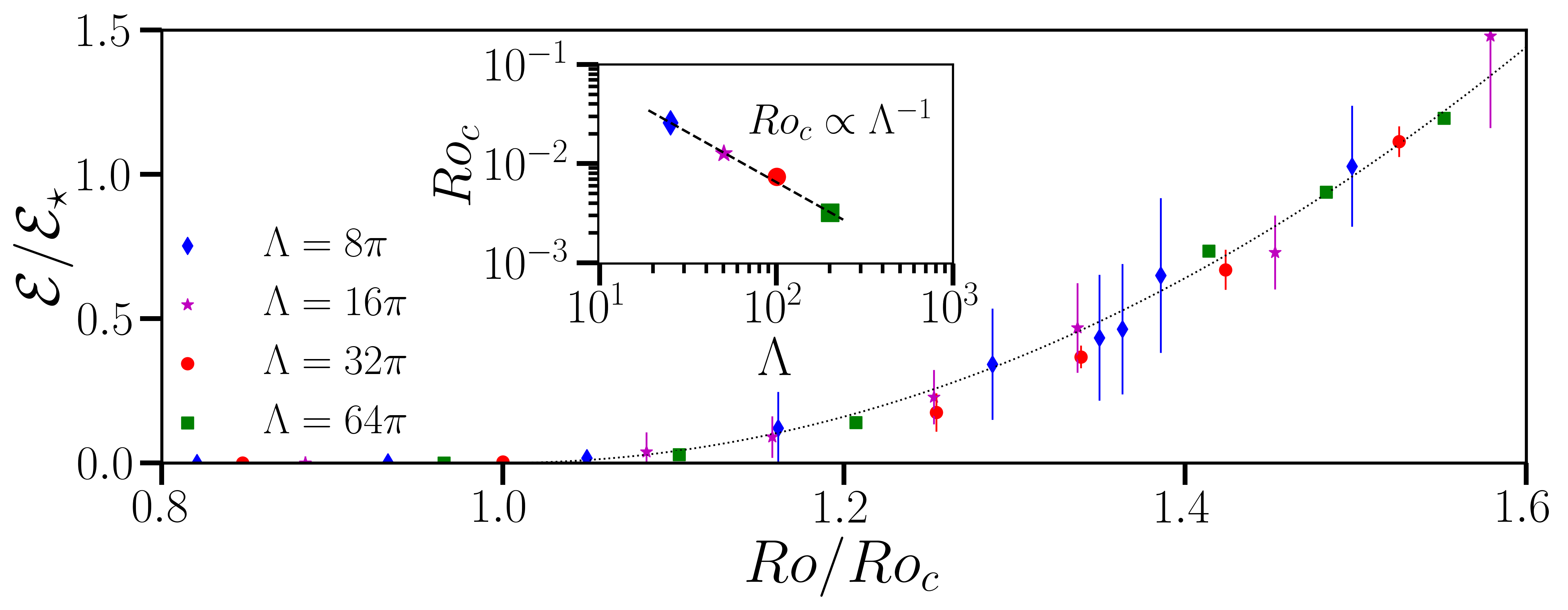} 
	\put(-210,80){{(b)}}
	\\
	\includegraphics[width=0.48\linewidth]{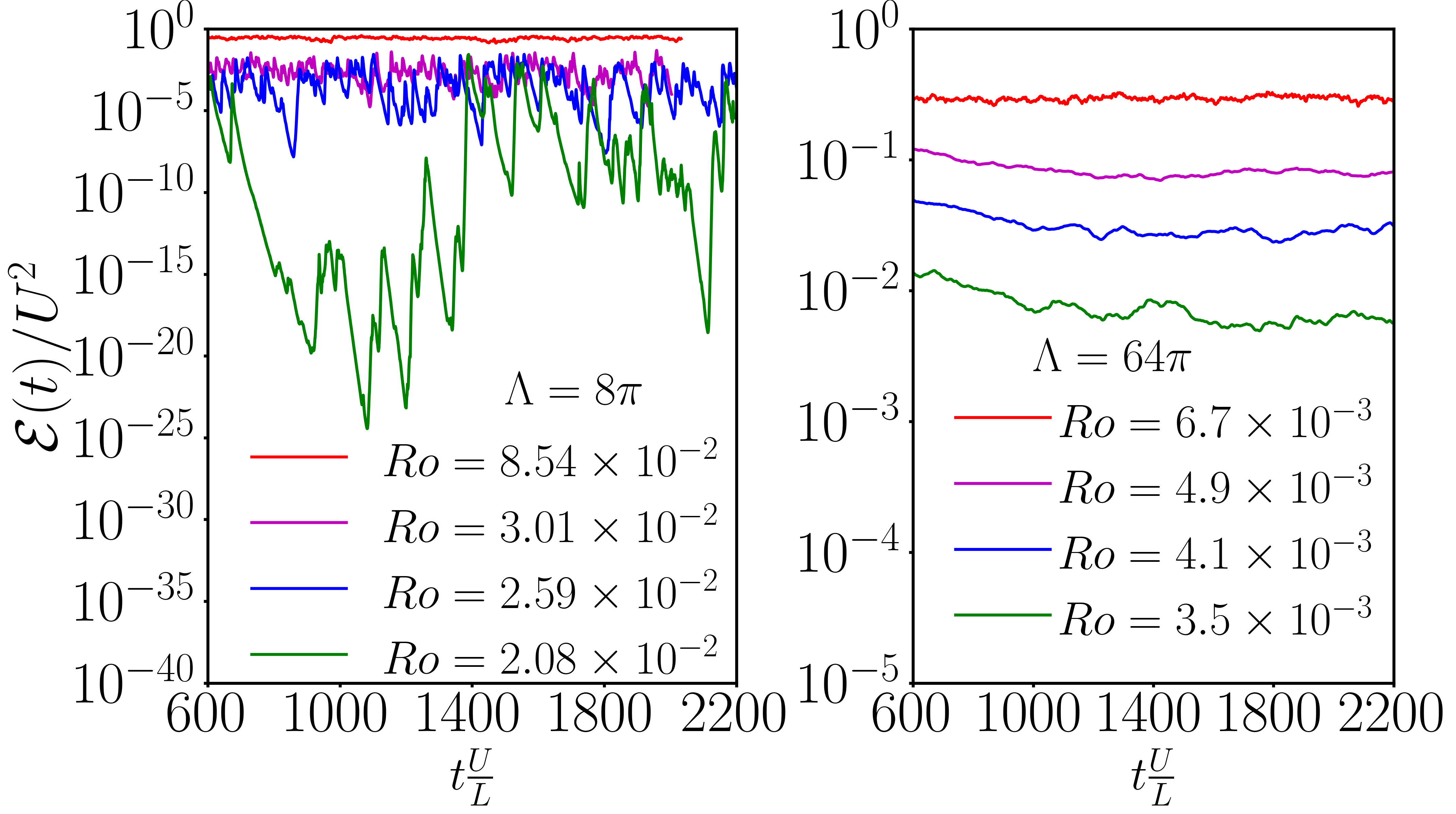}
	\put(-210,75){{(c)}}
        \put(-90,75){{(d)}}
	\includegraphics[width=0.48\linewidth]{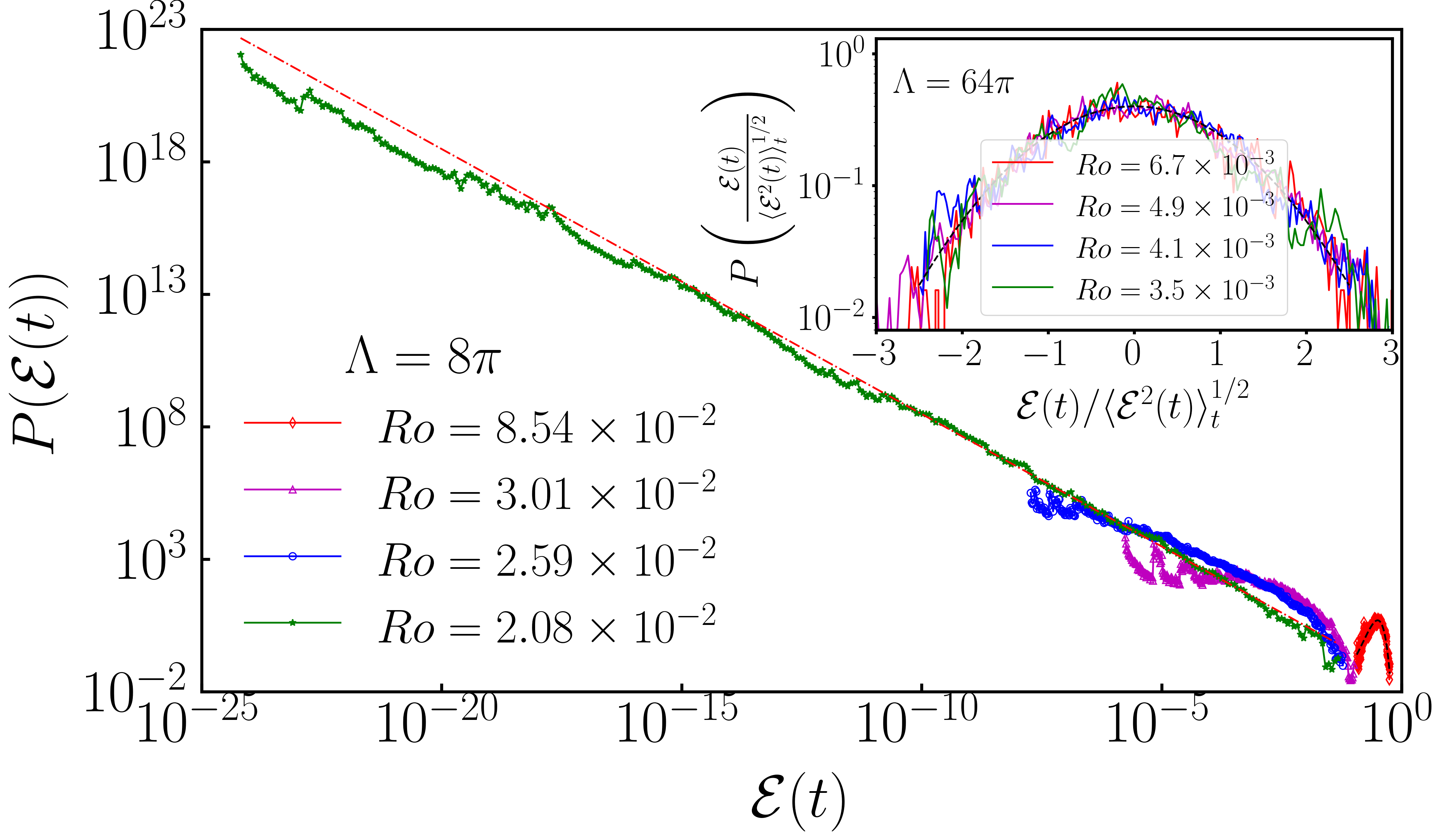}
	\put(-210,110){{(e)}}
	\caption{Order parameter and the bifurcation over a turbulent background. (a) Log-log plots of the mean 3D perturbation energy $\mathcal{E}/U^2$ vs $\Ro$ for different values of $\Lambda$. The shaded regions show the threshold range over which the bifurcation occurs as obtained from LSA. (b) Plots of $\mathcal{E}/\mathcal{E}_{\star}$ vs $\Ro/\Ro_c$ near the transition. Black dashed-line shows scaling with the critical exponent ($\beta$) observed with $\beta =  2$. Inset: Variation of $\Ro_c$ with $\Lambda$; black dashed line indicates the $\Lambda^{-1}$ scaling. (c) Time series of $\mathcal{E}(t)/U^2$ for different values of $\Ro$ for $\Lambda=8\pi$; (d) shows these plots for $\Lambda=64\pi$. (e) PDFs of $\mathcal{E}(t)$ for different $\Ro$ for $\Lambda=8\pi$ (the inset is for $\Lambda=64\pi$); the dashed black line in the inset indicates a Gaussian fit.}
	\label{fig:pertenergy}
\end{figure*}

\section{Theoretical framework}
\label{sec:theoframe}

\subsection{Governing equations and parameters}
We consider an incompressible fluid flow subjected to global rotation along the vertical axis, $\mathbf{e}_z$, in a triply periodic domain of size $L \times L \times H$, where $L$ is the horizontal length scale and $H$ is the depth of the domain along the $z$-direction. The velocity field, $\mbfu(\mbfx,t)$, in a frame rotating with speed $\Omega \mathbf{e}_z$, is governed by the Navier- Stokes equations:
\begin{equation}\label{eq:NSE}
 \frac{\partial \mbfu}{\partial t} + \mbfu \cdot \nabla \mbfu = -\frac{1}{\rho} \nabla p + 2 \Omega \mbfu \times \mbfe_z - \mu \mbfu + \nu \nabla^2 \mbfu + \mbff.
\end{equation}
Here $\mbfu$ is the velocity field obeying $\nabla \cdot \mbfu = 0$, $p(\mbfx,t)$ is the pressure field, $\rho$ the density, $\mu$ is the large-scale friction coefficient, and $\nu$ is the viscosity. The fluid flow is maintained in a statistically steady turbulent state by applying an external Kolmogorov force $\mbff(\mbfx) = f_0 \, \sin (k_f y) \, \mbfe_x$, $f_0$ being the forcing amplitude and $k_f$ is the forcing wave number; the chosen external force field is invariant along the rotation axis. Also, we set the aspect ratio of the domain $H/L=1/4$ for all our runs.

We can characterize the state of a rotating turbulent flow with the help of the following dimensionless parameters. The two Reynolds numbers based upon viscosity and friction coefficient are: $\Rey = UL/\nu$ and $\Rh = U/(\mu L)$,
where $U= {\langle |\mbfu|^2 \rangle}^{1/2}_{\mbfx,t}$ is the root-mean squared velocity and ${\langle \rangle}_{\mbfx,t}$ denotes averaging over both the space and time domains. The non-dimensional parameter $\Lambda = k_f L$  controls the scale separation between the box size ($L$) and forcing length scales. 
A measure of the relative strength of the inertial force to the Coriolis force is given by a dimensionless number, called the Rossby number: $\Ro = U/(2\Omega L)$

In general, the 2D solutions obtained by balancing the forcing and dissipation are unstable to 3D perturbations in the absence of rotation; they transition to a fully developed 3D turbulence if the $\Rey$ is sufficiently large. However, rotation and confinement stabilise the 2D flow. The 2D dimensional nature holds only up to an aspect-ratio-dependent critical $\Ro$, for a given $\Rey$, above which the three dimensional character emerges spontaneously~\cite{seshasayanan2020onset,lohani2024effect}.

Fast rotating turbulence involves processes spanning a multitude of length and time scales; the fastest time scales are associated with the inertial waves that obey the dispersion relation $\omega_{\pm}(\mathbf k) = \, \pm 2\Omega \, k_z/|\mathbf k|$, where $\mbfk$ is the wave vector and $k_z=\mathbf{e}_z\cdot \mbfk$. 
The transition from 2D to 3D flow regime, when $\Ro$ is varied, can be examined by splitting the velocity field into two components: The geostrophic mode $\mbfu_{2D}(x,y,t)$ (with $k_z=0$ in Fourier space) which evolves in the slow manifold ($\omega_{\pm}(\mathbf k) = 0$) and a perturbation velocity field $\widetilde{\mbfu}(\mbfx,t)$, evolving in the fast manifold ($\omega_{\pm}(\mathbf k) \ne 0$). The geostrophic mode is obtained by averaging the velocity field along the direction of rotation:
\begin{equation}\label{eqn:geomode}
	\mbfu_{2D}(x,y,t) = \dfrac{1}{H}\int^{H}_{0} \mbfu(\mbfx,t)\, dz.
\end{equation}
The perturbation velocity field is then simply given by 
\begin{equation}\label{eq:pertfield}
	\widetilde{\mbfu}(\mbfx,t) = \mbfu(\mbfx,t) - \mbfu_{2D}(x,y,t).
\end{equation}
Moreover, $ \mathcal{E}(t) = \langle \mid \widetilde{\mbfu}(\mbfx,t)\mid^2 \rangle_{\mbfx}$ provides a measure of the intensity of perturbations and can be used as an order parameter to characterize the transition, $\langle \cdot \rangle_{\mbfx}$ represents spatial averaging. The time averaged value is simply expressed as $\mathcal{E}= \langle \mathcal{E}(t) \rangle_t$.

In Fourier space, the total velocity field $\widehat {\mathbf{u}}(\mathbf{k})$ and the 3D perturbation velocity field $\widehat {\widetilde {\mathbf{u}}}(\mathbf{k})$ allow us to define the corresponding energy spectra as $E(k)=(1/2)\sum_{k\leq |\mathbf{k}|<k+1}|\widehat{\mathbf{u}}(\mathbf{k})|^2$ and  $\widetilde{E}(k)=(1/2)\sum_{k_z\neq 0, k \leq |\mathbf{k}|<k+1} |\widehat{\widetilde{\mathbf{u}}}(\mathbf{k})|^2$, respectively.

The LSA, around the 2D turbulent base flow, allows us to develop a simplified understanding of this transition from 2D to quasi-2D flow regime. It has been observed that the most unstable mode corresponds to the gravest vertical wave number, $q=2\pi/H$; see Refs.~\cite{seshasayanan2020onset,lohani2024effect} for a more detailed discussion. Here, we perform LSA over a domain of size $L \times L$ with $q=2\pi/H$ using Kolmogorov forcing at wave number $k_f$. The threshold thus obtained will be used to compare with 3D nonlinear solutions.

\subsection{Numerical simulations}

We perform DNSs of the Eq.~\eqref{eq:NSE} by using a pseudo-spectral method and a $2/3$ de-aliasing rule. For this purpose, we use GHOST~\cite{mininni2011hybrid}, a pseudo-spectral solver. In this study, we keep $\Rey$ and $\Rh$ fixed, up to a set tolerance  ($\leq 6\%$), by adjusting the relevant dissipation coefficients. Therefore, effectively we have only two free control parameters: $\Lambda$ and $\Ro$. We use the Taylor-Green velocity field as the initial data and perform each run for over few thousand eddy turnover time, after the steady-state is reached. In table~\ref{tab:runsD}, we provide the details of our DNS runs. 

\begin{table}
	\begin{center}
	\caption{\small Table of DNS runs. $N_x$, $N_y$ and $N_z$ are the number of collocation points along x, y and z, directions respectively. $N_z$ is chosen in order to have an isotropic grid spacing. For each set of DNS runs, Rossby number $\Ro$ is varied, while keeping $\Rey$ ($\mathcal{O}(10^4)$) and $\Rh$ ($\mathcal{O}(10^0)$) nearly fixed.}
	
	\label{tab:runsD}
		
		\begin{tabular}{l  c c  c  c  c } 
			\hline\hline
			{\tt Run}  & $\Rey$ & $\Rh$ & $\Lambda$ & $N_x=N_y$ & $N_z$ \\
			\hline
			{\tt Set1} & $1.1\times 10^{4}$ & $4.2$ & $8\pi$ & $256$ & $64$\\
			{\tt Set2} & $2.0\times 10^{4}$ & $3.5$ & $16\pi$ & $512$ & $128$\\
			{\tt Set3} & $1.6\times 10^{4}$ & $2.5$ & $32\pi$ & $512$ & $128$\\
			{\tt Set4} & $1.6\times 10^{4}$ & $2.0$ & $64\pi$ & $512$ & $128$\\
		\end{tabular}
	\end{center}
	
\end{table}

\section{Results}
\label{sec:results}

\subsection{Order parameter, bifurcations, and on-off intermittency}
We characterize the behaviour of the perturbation energy, $\mathcal{E}$ for different $\Ro$ and $\Lambda$. In Fig.~\ref{fig:pertenergy} (a) we show the normalized perturbation energy, $\mathcal{E}/U^2$, versus $\Ro$ for different values of $\Lambda$. Mean values of relevant quantities are calculated after reaching statistically steady state and averaging over few thousand eddy turn-over times. Blue, magenta, red, and green dotted curves show results corresponding to $\Lambda = 8 \pi$, $16 \pi$, $32 \pi$,  and $64 \pi$ sets respectively.

The limit $\Ro \to \infty$, at sufficiently large Reynolds numbers, corresponds to a flow where the effect of rotation is negligible, dominated by small-scale structures, while a 2D behaviour occurs at $\Ro \to 0$~\cite{gallet2015exact} and in between these two limits, the state is referred as quasi-2D state. In accordance with this, we observe that at large $\Ro$, the perturbation energy for $\Lambda=8\pi$ (blue line with `$\diamond$') saturates to a finite value. As we decrease $\Ro$ (or increase the rotation rate), the perturbation energy decreases sharply to zero for $\Ro \lesssim \Ro_c$. 
We take $\Ro_c$ as the Rossby number at which the perturbation energy $\mathcal{E}$ becomes nearly $0.01 \%$ of total energy ($U^2$). This behaviour indicates that below $\Ro_c$, the perturbations in the velocity field are strongly suppressed and the flow prefers to remain in the 2D state. We observe qualitatively similar behaviour for other values of $\Lambda=16\pi$, $32\pi$ and $64\pi$. However, the threshold $\Ro_c$ decreases with increase in $\Lambda$, as the 2D character appears at increasingly higher rotation rates. In Fig.~\ref{fig:pertenergy} (a), we also show the threshold range obtained from the LSA for each $\Lambda$, indicated by a vertical shaded region. The threshold $\Ro_c$ inferred from the DNS runs lie within the LSA range, thereby indicating a very good agreement between the two. The LSA is done on turbulent background and due to intermittent nature of perturbation growth in LSA and due to computational constraints one can determine accurately only the upper and lower threshold for the onset of the instability. Thus the threshold is expected to lie within the two values marked in different colors for different $\Lambda$. 3D solutions also have uncertainity due to the presence of intermittency near the threshold, though due to the nonlinear saturation we fix the total time of evolution and analyse the transition diagram from the averages.

This  behaviour of $\mathcal{E}/U^2$ with $\Ro$ shows that it can serve as an order parameter that saturates to a finite value for $\Ro \gg \Ro_c$ and is effectively zero for $\Ro < \Ro_c$; thus, it can distinguish between the two states. Moreover, the behaviour in Fig.~\ref{fig:pertenergy} (a) suggests that the transition is smooth. In order to characterize the bifurcation, in Fig.~\ref{fig:pertenergy} (b) we plot the normalised perturbation energy, $\mathcal{E}/\mathcal{E}_{\star}$ for different values of $\Lambda$ with $\Ro/\Ro_c$, where $\mathcal{E}_{\star}$ = $\mathcal{E}(\Ro = 1.5 \times \Ro_c)$. The vertical error bars represent the standard deviation of mean values in the steady state. Near $\Ro/\Ro_c \sim 1 $, it begins to slowly rise and we find that it can be fit to a form $(\Ro -\Ro_c)^{\beta}$, with $\beta \approx 2.0$, for all the values of $\Lambda$, indicated by a black dashed line in Fig.~\ref{fig:pertenergy} (b). We find that the accurate determination of the exponent $\beta$ is difficult, as the order parameter fluctuates strongly near the threshold and the exponent found from the fit is sensitive to the exact value of the threshold. 

The inset of Fig.~\ref{fig:pertenergy} (b) shows that $\Ro_c \sim \Lambda^{-1}$. This is in agreement with previous observations at high rotation rates~\cite{van2020critical} and was also conjectured in Ref.~\cite{alexakis2018cascades}, but for 3D to 2D transition. The transfer of energy towards 3D modes is possible only when the frequency of perturbations is slow enough to interact with the slow manifold. When this happens, we can equate the inertial frequency $2\Omega /(k_f H)$ with the one associated with non-linear time scales $U/L$: $2\Omega /(k_f H) \approx U/L$, which yields the scaling $\Ro_c \sim \Lambda^{-1}$, where $k_z  \sim 1/H$ is the most unstable wave number in the LSA and  $|\mathbf k| \sim k_f$ represents the typical size of vortices on which the instability develops. 

In order to gain further insights into the nature of dynamics, we analyse the time-series of the perturbation energies. Figure~\ref{fig:pertenergy} (c) shows the time evolution of $\mathcal{E}(t)/U^2$ for different rotation rates ($ 10^{-2} \lesssim \Ro \lesssim 10^{-1}$) at fixed $\Lambda = 8\pi$. For $\Ro=8.54 \times 10^{-2} > \Ro_c$ (red solid line), the fluctuations are small. However, a comparison of this with the magenta curve ($\Ro=3.01\times 10^{-2}$) and the blue curve ($\Ro=2.59\times 10^{-2}$) shows that at lower values of Rossby numbers (but $\Ro \gtrsim \Ro_c$), the mean decreases and the fluctuations start to become large. Moreover, as we approach close to $\Ro_c$ from above, the fluctuations become more and more intermittent; see the green curve for $\Ro=2.08\times 10^{-2}$. In Fig.~\ref{fig:pertenergy} (d) we show $\mathcal{E}(t)/U^2$ for $\Lambda=64\pi$ for $\Ro=6.7\times 10^{-3}$ (red solid line), $\Ro=4.9\times 10^{-3}$ (magenta solid line), $\Ro=4.1\times 10^{-3}$ (blue solid line), and $\Ro=3.5 \times 10^{-3}$ (green solid line). It clearly shows that the extent of fluctuations is small compared to the case of $\Lambda=8\pi$. Also, unlike the previous case, the temporal behaviour lacks intermittent character for Rossby numbers close to $\Ro_c$. For other $\Lambda$ values, we observe intermittent behaviour similar to $\Lambda=8 \pi$ case, but intermittent bursts become less prominent as $\Lambda$ is increased, and finally for $\Lambda = 64 \pi$, these are absent. This clearly indicates that the region of intermittency can be sensitive to the value of $\Lambda$ and $\Ro$.

We characterize the temporal fluctuations of the order parameter by computing the probability distribution function (PDF) of $\mathcal{E}(t)$. In Fig.~\ref{fig:pertenergy} (e) and its inset we show the PDFs for different values of $\Ro$ for the two values of $\Lambda=8\pi$ and $64\pi$, respectively, and directly correspond to the two cases discussed above in Figure~\ref{fig:pertenergy} (c) and (d).  For the small rotation rates (large $\Ro$), the PDFs are Gaussian, e.g., the PDF for $\Ro=8.54 \times 10^{-2}$ and $\Lambda=8\pi$ (red line with `$\diamond$'); black dashed line indicates the Gaussian fit. When we decrease $\Ro$ ($\sim \mathcal{O}(10^{-2})$), while keeping $\Lambda=8\pi$ fixed, the nature of the PDFs changes, we see an emergence of a power-law region, for smaller values of the perturbation energies, $\mathcal{E}(t)$. A comparison of PDFs in Fig.~\ref{fig:pertenergy} (e) for $\Ro=3.01 \times 10^{-2}$ (magenta line with `$\triangle$'), $2.59 \times 10^{-2}$ (blue line with `$\circ$'), and $2.08 \times 10^{-2}$ (green line with `$\star$') show that the extent of the power-law region $P(x) \sim x^{-\alpha}$ significantly increases, as $\Ro \to \Ro_c$. In fact, it extends over eight decades of $\mathcal{E}(t)$ for $\Ro = 2.08 \times 10^{-2}$ ($ \simeq \Ro_c$); the red dashed-dotted line indicates that the exponent $\alpha \approx 1$. A similar behavior is observed for the $\tt Set2$ and $\tt Set3$  with $\Lambda=16\pi$ and $32\pi$, respectively, as well. However, for the $\tt Set4$ with $\Lambda=64\pi$, we observe that the PDFs are Gaussian for Rossby numbers down to $\Ro_c$, as we show in the inset of Fig.~\ref{fig:pertenergy} (e) for $\Ro=6.7 \times 10^{-3}$ (red line), $\Ro=4.9 \times 10^{-3}$ (magenta line), $\Ro=4.1 \times 10^{-3}$ (blue line ), and $\Ro=3.5 \times 10^{-3}$  (green line); the black dashed curve in the inset shows a Gaussian curve for comparison. 

\subsection{Absence of hysteresis}

\begin{figure}
	\centering
	\includegraphics[width=0.48\linewidth]{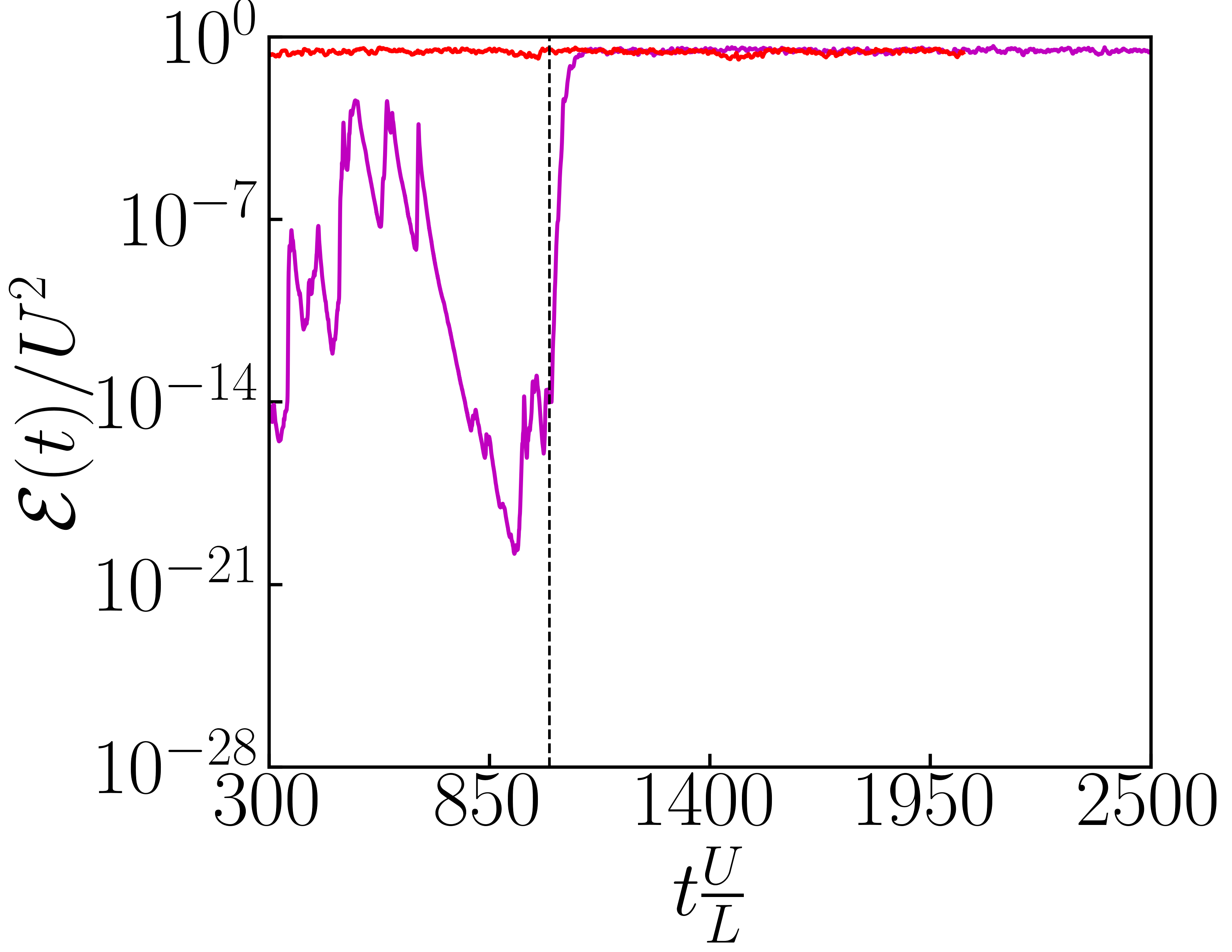}
	\put(-90,25){{(a)}}
	\includegraphics[width=0.50\linewidth]{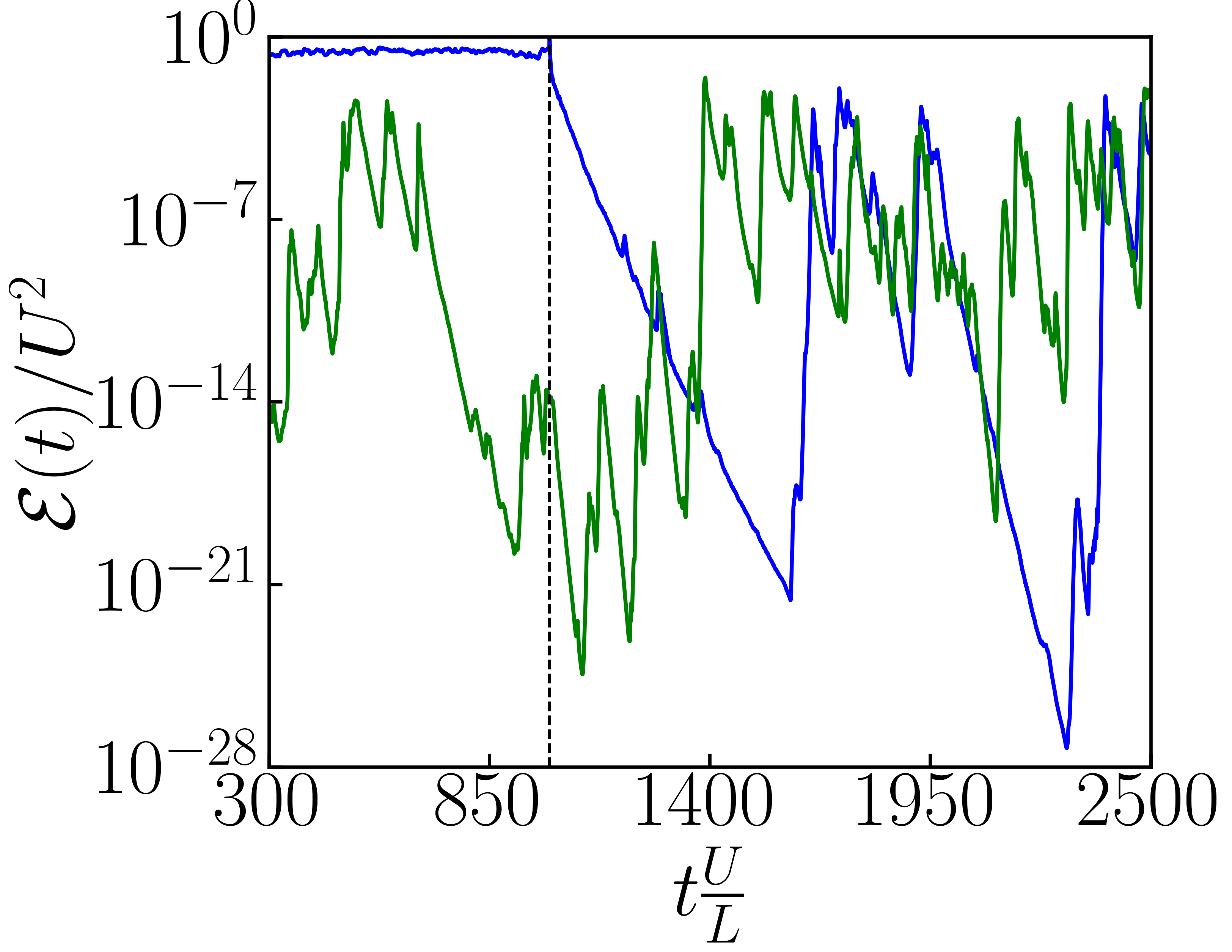} 
	\put(-94,25){{(b)}}
	\caption{Time series of the 3D perturbation energy $\mathcal{E}(t)/U^2$ at $\Lambda=8\pi$ (a) solid red color curve refers to the time series at fixed $\Ro=8.54\times 10^{-2}$, time series shown using magenta color corresponds to protocol $\tt P1$; (b) solid green color curve shows the time series for fixed $\Ro=2.08\times 10^{-2}$, blue solid curve refers to protocol $\tt P2$. The vertical black dashed line indicates the time at which $\Ro$ is changed.}
	\label{fig:hysteresis}
\end{figure}

\begin{figure*}
	\centering
	\includegraphics[width=0.95\linewidth]{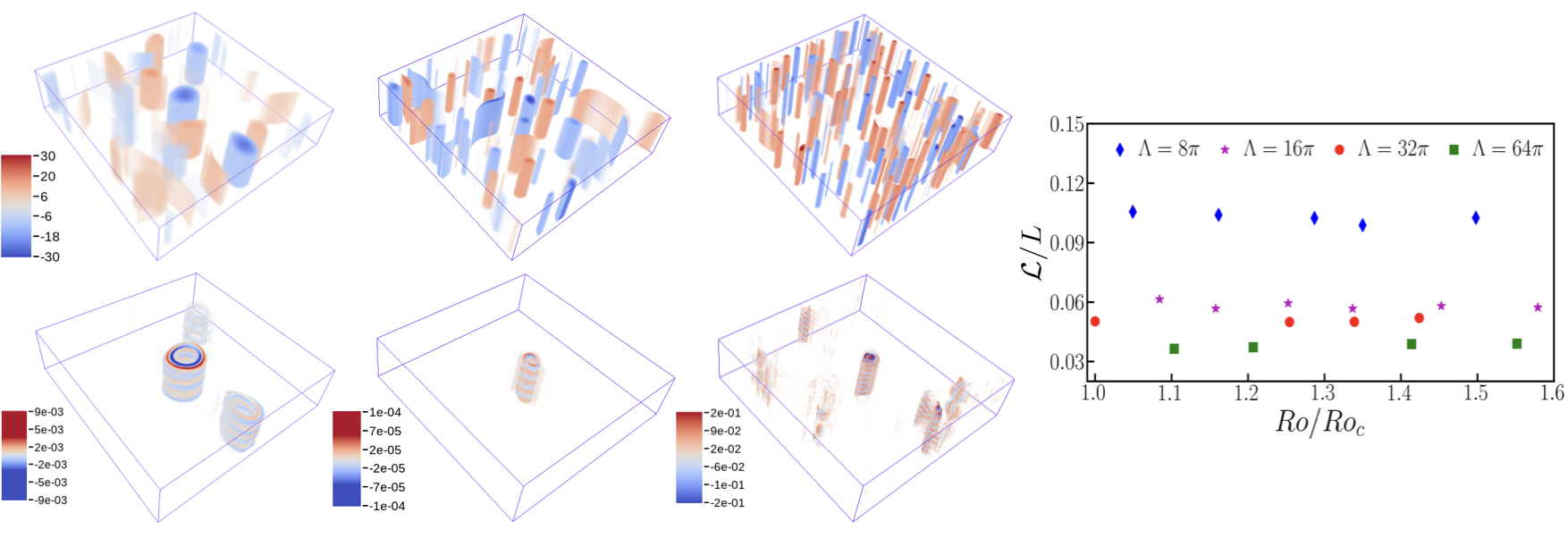}
	\put(-495,155){\small{(a)}}
	\put(-480,150){\small{(a.1)}}
	\put(-380,150){\small{(a.2)}}
	\put(-280,150){\small{(a.3)}}
	\put(-480,75){\small{(a.4)}}
	\put(-380,75){\small{(a.5)}}
	\put(-280,75){\small{(a.6)}}
	\put(-170,155){\small{(b)}}

	\caption{Centrifugal instability driving the bifurcation. (a) [Top panel] Isosurface plots of the vorticity field $\omega_{2D}$: (a1) $\Lambda = 8\pi$, (a2) $\Lambda=16\pi$ (a3) $32\pi$. [Lower panel] Isosurface plots of the 3D perturbation vorticity field $\widetilde{\omega}_z$: (a4) $\Lambda = 8\pi$, (a5) $\Lambda=16\pi$ (a6) $32\pi$. All these figures have $\Ro \gtrsim \Ro_c$. (b) Length scale $\mathcal{L}/L$ associated with the instability vs $\Ro/\Ro_c$ for $\Lambda=8\pi$ (blue $\diamond$ markers), $16\pi$ (magenta $\star$ markers), $32\pi$ (red $\circ$ markers), and $64 \pi$ (green $\square$ markers).}
	\label{fig:instmechvort}
\end{figure*}
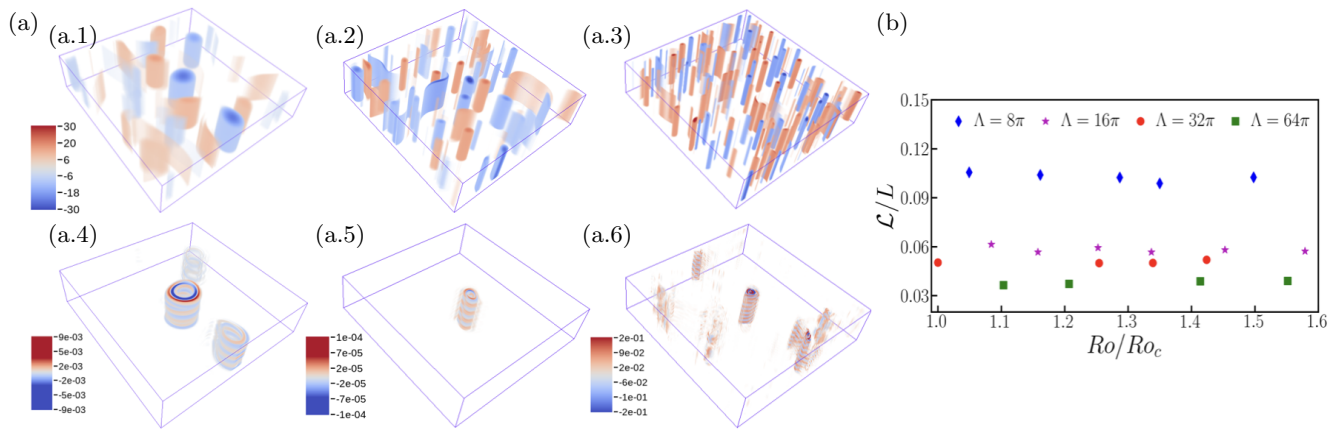

In Fig.~\ref{fig:hysteresis} (a) we show the behavior of the 3D perturbation energy, $\mathcal{E}(t)/U^2$, time series, when the Rossby number is abruptly increased from $\Ro=2.1\times 10^{-2}$ (magenta solid line, state $\tt S_1$) to $8.54\times 10^{-2}$ (state $\tt S_2$) for $\Lambda=8\pi$ at $tU/L \approx 1000$ indicated by the black dotted line. We refer to this as protocol $\tt P1$. For comparison, we also show the steady state energy time series from the run with $\Ro=8.54\times 10^{-2}$ (red solid line, state $\tt S_3$) where $\Ro$ is fixed. We find that the energy time series obtained, after the abrupt change in the Rossby number to $\Ro=8.54\times 10^{-2}$ (magenta solid line, $\tt S_2$), agrees very well with that of a run with constant $\Ro=8.54\times 10^{-2}$ (red solid line, $\tt S_3$) for $tU/L > 1000$ (beyond black dashed line). Similarly, in Fig.~\ref{fig:hysteresis} (b), we demonstrate the result of the reverse protocol $\tt P2$, in which $\Ro$ is abruptly decreased from $8.54\times 10^{-2}$ to $2.1 \times 10^{-2}$ for $\Lambda=8\pi$ at $tU/L \approx 1000$ (see the behaviour of the blue solid line). Once again, we observe an excellent agreement between the newly adjusted state (blue solid line, state $\tt S_4$) and the steady state at $\Ro=2.1\times 10^{-2}$ (green solid line, $\tt S_1$).

In summary, the protocol $\tt P1$ takes the system from a state near the threshold in the quasi-2D flow regime to another state far from it in the same regime, whereas the protocol $\tt P2$ implements the reverse transition. The equivalence of the states $\tt S_2 \equiv S_3$ and $\tt S_4 \equiv S_1$, after the change in $\Ro$, indicates the absence of hysteresis. Furthermore, we do not observe hysteresis when the control parameter is changed from a value below the critical point to a value above it and vice versa (not shown).

\subsection{Instability mechanisms near the bifurcation}

To better understand the flow near the bifurcation threshold, we examine the structures that are, in particular, associated with the underlying instability mechanisms. In Fig.~\ref{fig:instmechvort} (a.1), (a.2) and (a.3), we plot the isosurfaces of the vorticity field $\omega_{2D}$($=\mathbf{e}_z \cdot \nabla \times \mbfu_{2D}$)  for $\Lambda=8\pi$, $16\pi$ and $32\pi$, respectively, for $\Ro \simeq \Ro_c$, whereas in the lower panel, Fig.~\ref{fig:instmechvort} (a.4), (a.5) and (a.5), we show the corresponding perturbation vorticity field $\widetilde{\omega}_z$ ($=\mathbf{e}_z\cdot\nabla \times \widetilde{\mbfu}$). Note that these snapshots are from a period of instantaneous growth of the perturbation energy. 

It is interesting to observe from the lower panel that the perturbation vorticity field exhibits intense flow structures near the contra-rotating vortices in the top panel. Each flow structure is strongly localized around the contra- rotating vortices in the form of a helical coil with alternating positive and negative vorticity along the axis of rotation. The 2D projection is consistent with azimuthal mode $m=1$, as reported in Ref.~\cite{lohani2024effect}. For $\Lambda = 64 \pi$, spatial structures being comparatively smaller, have not been shown; the corresponding LSA shows centrifugal instability for this case as well.  

For the Reynolds numbers ($\Rey$ and $\Rh$) and the range of explored $\Lambda$ and $\Ro$, we do not find any clear evidence to suggest the presence of parametric instability. This is in agreement with the results of Ref.~\cite{lohani2024effect}, wherein the study of a single mode 3D perturbation coupled to 2D geostrophic modes showed that the centrifugal instability is present at moderate $\Rey$ and $\Ro$, while at large $\Rey$ and low $\Ro$ parametric instability is found. However, note that the parametric instability can be the destabilization mechanism at similar or even low $\Rey$ and moderate $\Ro$ for a sufficiently elongated domain~\cite{lohani2024effect}. 
 
Finally, we determine the horizontal length scale $\mathcal{L}=\langle \int \ell_{\perp} \tilde{E}(\mbfk)\, d\mbfk/\int \tilde{E}(\mbfk)\, d\mbfk \rangle_t$ associated with instability as the weighted average of $\ell_{\perp}=1/\sqrt{k^2_x+k^2_y}$, with $\widetilde{E}(\mbfk)$ as weight function, where $k_x, k_y$ are the wave vector components perpendicular to the axis of rotation. Figure~\ref{fig:instmechvort} (b) shows that $\mathcal{L}/L$ remains independent of $\Ro$, but varies with $\Lambda$. This is again consistent with the instability mechanism described by the LSA.

\subsection{Spectral characterization: Energy spectra and fluxes}

\begin{figure*}[t]
	\centering
	\includegraphics[width=0.48\linewidth]{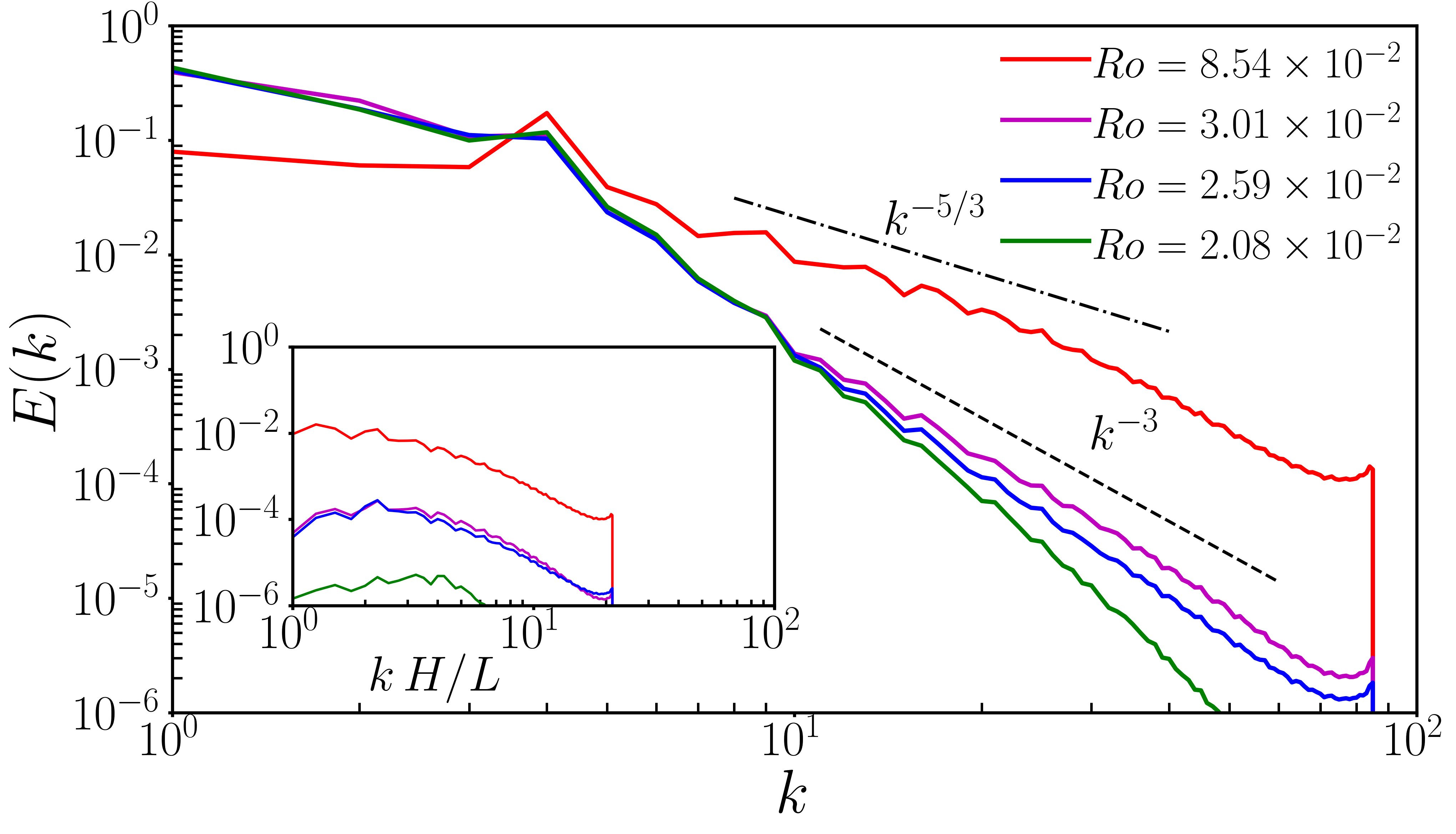}
	\put(-210,100){\small{(a)}}
	\includegraphics[width=0.48\linewidth]{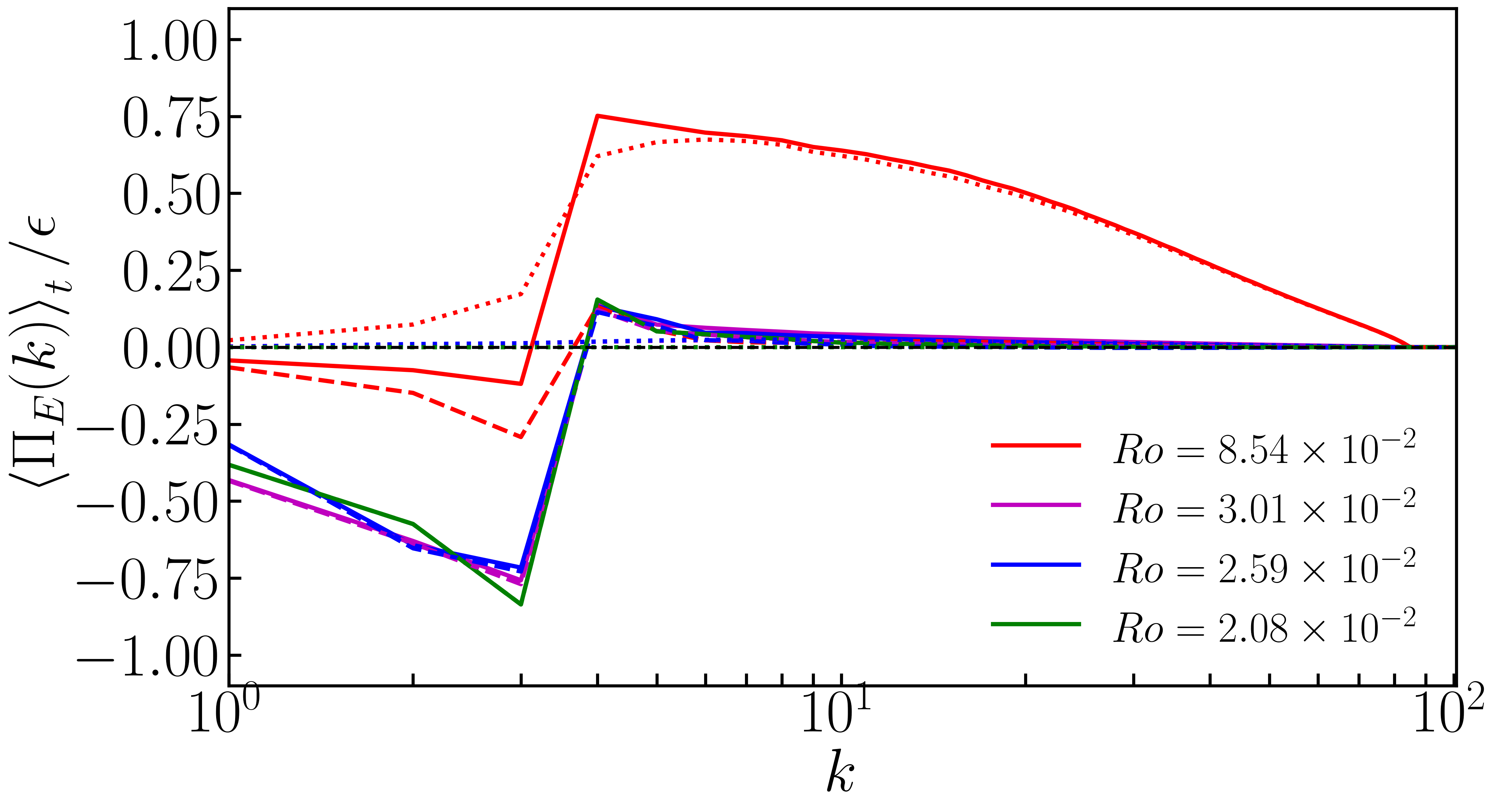}
	\put(-30,120){(b)}\\
	\includegraphics[width=0.48\linewidth]{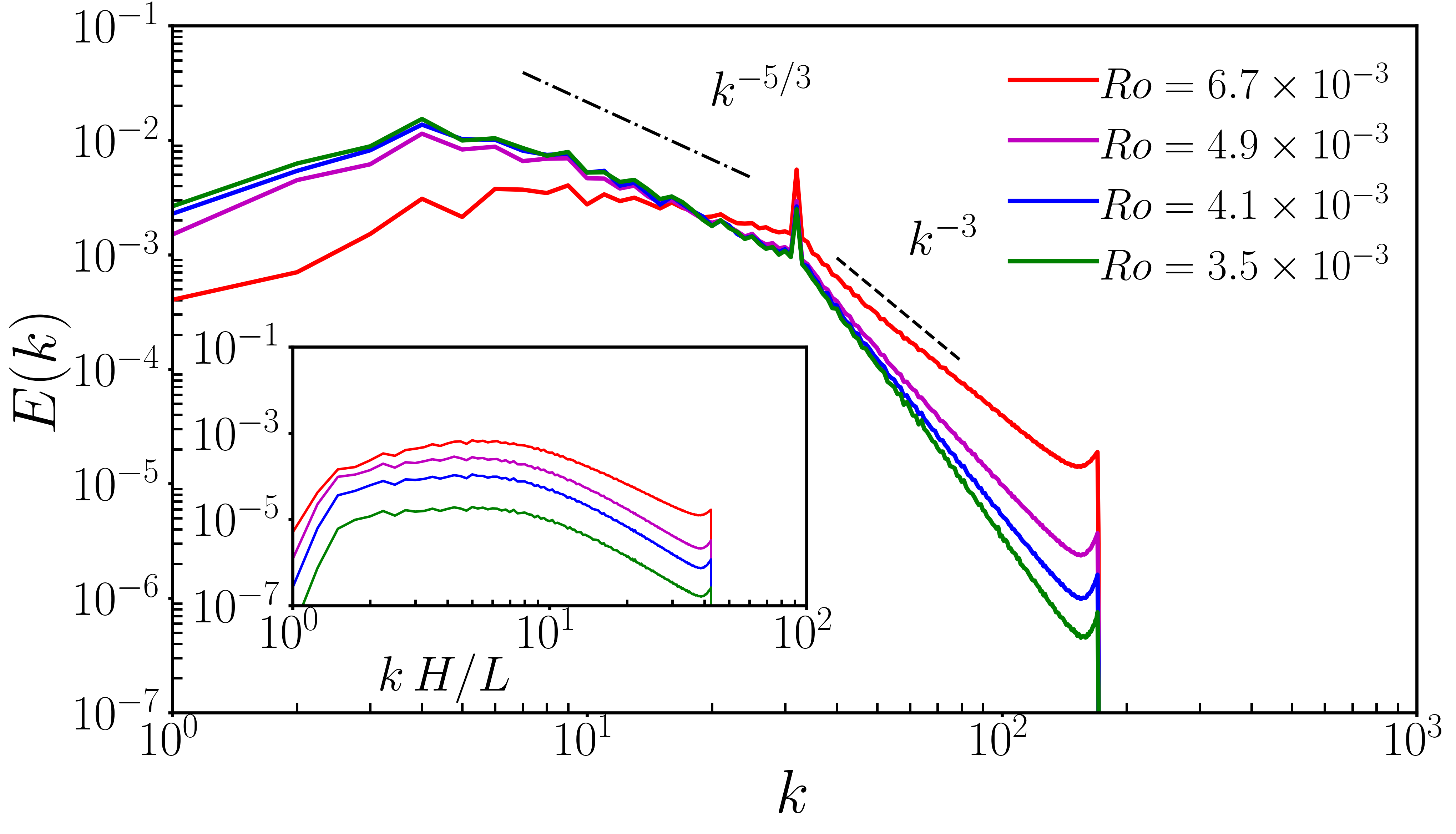}
	\put(-210,120){(c)}
	\includegraphics[width=0.48\linewidth]{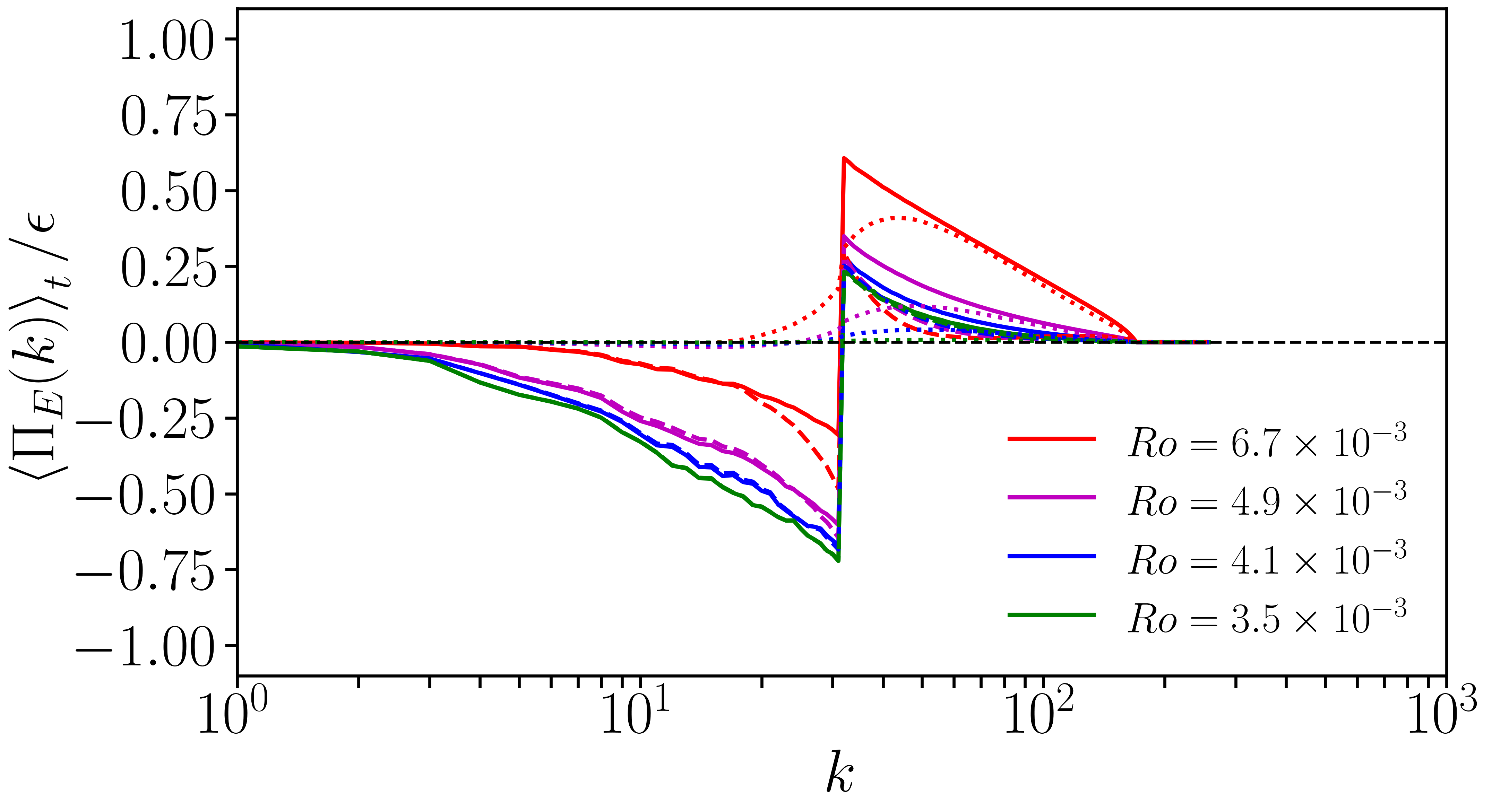}
	\put(-30,120){(d)}
	\caption{Energy spectra and fluxes: [Top panel]  $\Lambda=8\pi$: (a) Total energy spectra, $E(k)$, and inset shows 3D perturbation energy spectra, $\widetilde{E}(k)$, (b) Normalised energy flux $\Pi_{E}(k)$, at Rossby numbers $\Ro=8.5\times 10^{-2}$ (red solid line), $3.0\times 10^{-2}$ (magenta solid line), $2.59\times 10^{-2}$ (blue solid line), and $2.08\times 10^{-2}$ (green solid line). [Lower panel] $\Lambda=64\pi$: (c) $E(k)$, inset $\widetilde{E}(k)$, (d) Normalised energy flow $\Pi_E(k)$ at Rossby numbers $\Ro=6.7\times 10^{-3}$ (red solid line), $4.9\times 10^{-3}$ (magenta solid line), $4.1\times 10^{-3}$ (blue solid line), and $3.5\times 10^{-3}$ (green solid line). $\epsilon$ is the energy injection rate $\langle \mathbf{u}\cdot \mathbf{F}\rangle$. Dashed and dotted curves in (b) and (d) show the contribution from the 2D $(\Pi_{2D})$ and 3D $(\Pi_{3D})$ modes respectively.}
	\label{fig:spectquant}
\end{figure*} 

In Fig.~\ref{fig:spectquant} we analyse the spectral transport of energy under the combined effect of rotation and confinement. In all runs, the smallest non-zero vertical wave number $k_z$ is $4$, since $L/H=4$. We define and compute the total flux as $\Pi (k) = \Pi_{2D}(k) + \Pi_{3D}(k)$, where $\Pi_{2D}(k)$ represents the flux due to interactions purely among the 2D modes themselves and $\Pi_{3D}(k)$ is the flux contribution due to interactions that involve 3D perturbation velocity field. These are given by
\begin{subequations}\label{eq:flux_cal}
	\begin{align}
		\Pi_{2D}(k) &= -\langle \mathbf{\hat {u}}_{2D}^{<k} \cdot (\mathbf{\hat {u}}_{2D} \cdot \nabla \mathbf{\hat {u}}_{2D}) \rangle, \\
		\Pi_{3D}(k) &= -\langle \mathbf{\hat {\widetilde {u}}}^{<k} \cdot (\mathbf{\hat {\widetilde {u}}} \cdot \nabla \mathbf{\hat {\widetilde {u}}}) \rangle 
		- \langle \mathbf{\hat {u}}_{2D}^{<k} \cdot (\mathbf{\hat {\widetilde {u}}} \cdot \nabla \mathbf{\hat {\widetilde {u}}}) \rangle \nonumber \\
		& -\langle \mathbf{\hat {\widetilde {u}}}^{<k} \cdot (\mathbf{\hat {\widetilde {u}}} \cdot \nabla \mathbf{\hat {u}}_{2D}) \rangle
		-\langle \mathbf{\hat {\widetilde {u}}}^{<k} \cdot (\mathbf{\hat {u}}_{2D} \cdot \nabla \mathbf{\hat {\widetilde {u}}}) \rangle,
	\end{align}
\end{subequations}
where $\mbfu^{<k}$ represents a filtered velocity field with wave numbers less than $k$. The first term in $\Pi_{3D}$ (Eq.~\eqref{eq:flux_cal} (b)) represents the flux contribution due to interactions purely among 3D modes, whereas the remaining terms capture the transfer of energy due to the interaction of 3D modes with 2D modes.
  
Figure~\ref{fig:spectquant} (a) and (b) show the effect of rotation, on the energy spectra and the flux, respectively, for $\Lambda =8 \pi$. For $\Lambda=8\pi$ flow exhibits features that are a mix of 2D and 3D turbulence. This is evident from Fig.~\ref{fig:spectquant} (a), the energy spectra exhibits $k^{-5/3}$ for $k > k_f$, when the rotation rate is not large ($\Ro=8.5 \times 10^{-2}$); however, we find that at large rotation rates (small $\Ro$), the above scaling is replaced by a $k^{-\delta}$ region, with $\delta > 3$. The inset shows the perturbation energy spectra $\widetilde{E}(kH/L)$ is diminished at small $\Ro$. Also, at $\Ro \sim \mathcal{O}(10^{-2})$ it is insignificant compared to those from relatively slowly rotating domains at $\Ro \sim \mathcal{O}(10^{-1})$.

Figure~\ref{fig:spectquant} (b) shows the total energy flux $\Pi$ (solid line), the geostrophic mode contribution $\Pi_{2D}$ (dashed line), and the contribution due to 3D perturbations $\Pi_{3D}$ (dotted line) for $\Ro=8.54 \times 10^{-2}$ (red color), $3.0 \times 10^{-2}$ (magenta color), $2.59 \times 10^{-2}$ blue color, and $2.08\times 10^{-2}$ (green color).  At small $\Ro$, the energy fluxes become negative in the range $1<k < k_f$; hence, the energy is transfered to larger length scales, while having a small forward cascade in the range $k > k_f$. The anisotropy of the domain results in a dual cascade picture even for large $\Ro$, this is evident from a small inverse flux for $\Ro = 8.54 \times 10^{-2}$ (which is far from $\Ro_c$) in Fig.~\ref{fig:spectquant} (b).  At large $\Ro$, geostrophic modes lose some energy to 3D modes, the situation is akin to that of a well known dual cascade scenario in rotating turbulence. Figures~\ref{fig:spectquant} (c)-(d) show the dual cascade scenario for different $\Ro$ for $\Lambda=64\pi$.

\section{Conclusions and Discussion}
\label{sec:disccon}

We have examined and characterized the transition from 2D to quasi-2D flow regimes over a turbulent background, under the combined effect of rotation and confinement using 3D DNS. We find that the perturbation energy serves as a good candidate for the order parameter and exhibits a bifurcation with exponent $\beta \approx 2.0$, at a critical value of Rossby number ($\Ro_c$), when we vary $\Ro$ either from a very small to a large value or vice-versa. A clear theoretical understanding of these exponents, associated with transitions over turbulent background, is still lacking. However, studies of spatial and temporal correlations of fluctuations have helped in elucidating the origins of such exponents~\cite{van2021levy,alexakis2021symmetry}. Also, we find that hysteresis is absent and the transition is continuous. We emphasize that this 2D to quasi-2D transition is different from the abrupt transition from isotropic 3D to quasi-2D regimes that occurs at much higher Rossby numbers and exhibits hysteresis~\cite{yokoyama2017hysteretic}. To better understand these two transitions, the role of the nature of forcing mechanism and boundary conditions must be examined more closely~\cite{alexakis2015rotating,yokoyama2017hysteretic}.

 We find that the critical rotation rate varies linearly with the forcing wave number $k_f$, i.e., $\Ro_c \propto \Lambda^{-1}$. For the $\Rey$ considered in this study ($\Rey \simeq \mathcal{O}(10^4)$), the centrifugal instability is the primary mechanism that destabilizes the 2D flow regime and drives the system towards a 3D flow regime. The typical length scales associated with the instability are independent of $\Ro$, but vary with $\Lambda$. Moreover, near $\Ro_c$ the perturbation energy is highly intermittent for $\Lambda=8\pi$. This is further evident from the corresponding PDFs of perturbation energy fluctuations, which exhibit a power-law ($P(x) \simeq x^{-\alpha}$) with exponent $\alpha \simeq 1$. The value of the exponent is consistent with the phenomenon of on-off intermittency. However, the intermittency is less pronounced or absent at higher values of $\Lambda$. 

The spectral energy distribution reflects dual cascade scenario at large $\Ro$ and inverse cascade in the small $\Ro$ limit. The individual contributions from the geostrophic modes and perturbation modes show inverse cascade and direct cascade, respectively. The linear threshold obtained from a reduced model in Ref.~\cite{lohani2024effect}, which took into account only the dynamics of the most unstable 3D mode, is in very good agreement with our fully nonlinear DNS results. Furthermore, the reduced model and the DNS results together support that the LSA based approach is able to capture the transition from 2D turbulence to 3D flow regimes on a turbulent background at moderate Reynolds. A comprehensive understanding of these nonequilibrium transitions will require computation  of statistics over a much longer duration and a wide range of $\Rey$ values in both numerical and controlled experimental studies.

\acknowledgments
CSL acknowledges the MHRD, Govt. of India for the fellowship. SKN thanks the Prime Minister's Research Fellows (PMRF) scheme, Ministry of Education, Govt of India. This work used the Supercomputing facility of IIT Kharagpur established under the National Supercomputing Mission (NSM), Government of India and supported by CDAC, Pune. Authors acknowledge  the following NSM Grants DST/NSM/R\&D HPC Applications/2021/03.21 and DST/NSM/R\&D HPC Applications/2021/03.11.

\bibliography{reference}

\end{document}